# SimuLearn: Fast and Accurate Simulator to Support Morphing Materials Design and Workflows

Humphrey Yang[1], Kuanren Qian[2], Haolin Liu[2], Yuxuan Yu[2], Jianzhe Gu[1], Matthew McGehee[3], Yongjie Jessica Zhang[2], Lining Yao[1]

[1]HCI Institute,
Carnegie Mellon University
{hanliny, jianzheg, liningy}@andrew.cmu.edu

[2]Mechanical Engineering,
Carnegie Mellon University
{kuanrenq, haolinl, yuxuany1, jessicaz}@andrew.cmu.edu

[3]School of Design,
Carnegie Mellon University
mmcgehee@andrew.cmu.edu

**ABSTRACT**

Morphing materials allow us to create new modalities of interaction and fabrication by leveraging the materials' dynamic behaviors. Yet, despite the ongoing rapid growth of computational tools within this realm, current developments are bottlenecked by the lack of an effective simulation method. As a result, existing design tools must trade-off between speed and accuracy to support a real-time interactive design scenario. In response, we introduce SimuLearn, a data-driven method that combines finite element analysis and machine learning to create real-time (0.61 seconds) and truthful (97% accuracy) morphing material simulators. We use mesh-like 4D printed structures to contextualize this method and prototype design tools to exemplify the design workflows and spaces enabled by a fast and accurate simulation method. Situating this work among existing literature, we believe SimuLearn is a timely addition to the HCI CAD toolbox that can enable the proliferation of morphing materials.

**CSS Concepts**
• **Human-centered computing~Human computer interaction (HCI)**; *Interactive systems and tools; User interface programming.*

**INTRODUCTION**

In recent years, the HCI community has become interested in using morphing materials to enable new modes of interactions. These materials allow us to create shape-changing interfaces that are electricity-free and can respond to surrounding stimuli [37], the wearer's physiological conditions [46], or to realize novel fabrication methods [41]. However, due to their spatiotemporal behaviors and nonlinear material properties, it is difficult to predict the performances of morphing materials design. As a result, conventional computer-aided design (CAD) tools often have to make tradeoffs between speed and accuracy. In HCI, this complication further poses a challenge in making design tools because both real-time interactivity and visual fidelity are desired to inform design decisions.

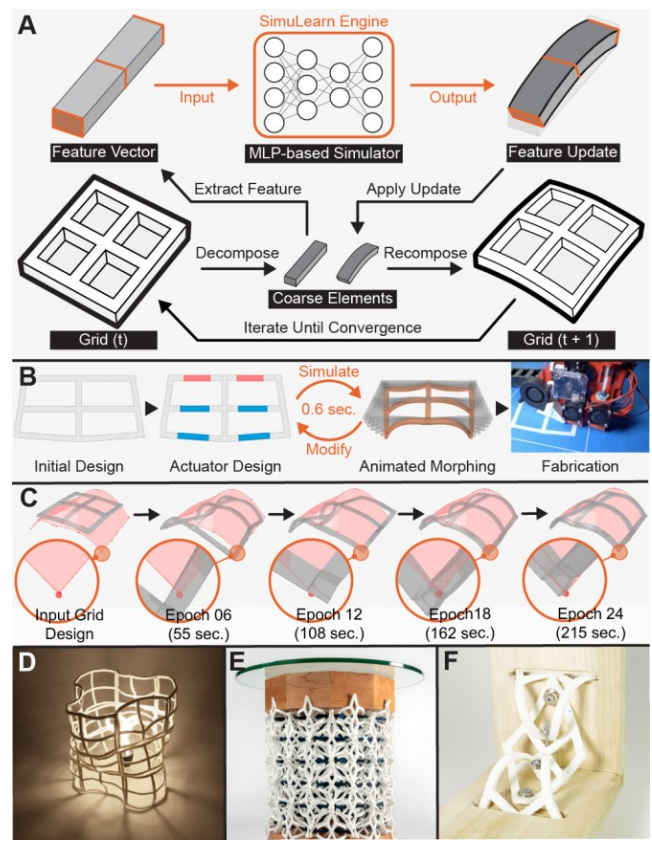

Figure 1. SimuLearn overview - (A) the computational theme of SimuLearn enables fast and high-fidelity (B) forward design iterations and (C) inverse design optimizations. These workflows enable design spaces that demand both simulation speed and accuracy, such as (D) modularization (lampshade), (E) material-driven parametric design (table stand), and (F) interlocking mechanisms (decorative joinery).

Existing simulation methods can be divided into three categories: geometrical methods, mass-spring models, and finite element analysis (FEA). Geometrical methods predict material performance by modeling the relationship between design parameters and experimental data (e.g., associating the length [1, 41] or layer thickness [40] of a printed thermoplastic actuator with its resulting bending angle). While they are fast to compute, these methods often take few if any physical parameters into account, and thus are not physically accurate. Alternatively, mass-spring models [12] seek to incorporate some physical factors present in the actuation environment, but they cannot account for the complex, nonlinear physics inherent to morphing materials and are prone to diverge. Advanced methods such as elastic rods [5,31] are also restricted to certain material properties and shapes, thus having a limited morphing materials design space. In contrast, while FEA is physically-based, their sheer computational cost renders them unviable in interactive design tools [47]. More, morphing materials are often soft during transformation and have virtually infinite degrees of freedom, requiring high-resolution discrete models to avoid divergence, which further slows down the computation. While model reduction methods [2,44] can be used to achieve interactive FEA, they require pre- and re-processing whenever the model geometry is changed. Therefore, they are less ideal for supporting iterative design workflows.

To address the need for an effective simulation method that allows an interactive design process of morphing materials, we propose SimuLearn, a data-driven simulation technique that combines FEA with machine learning (ML) to make physically accurate predictions in real-time (Figure 1C). This method takes FEA-generated data to ensure simulation accuracy and uses ML to generalize and achieve fast computation. W apply this concept to 4DMesh-like 2x2 grid structures [41] to demonstrate this simulation technique and showcase SimuLearn's workflow applicability. Results show that SimuLearn can produce high-quality simulations (97% accuracy) in real-time (0.61 seconds, over 1000 times faster than state-of-the-art FEA models). While the accuracy requirements may differ between use scenarios, we show that CAD tools based on this simulator (Figure 1A) can readily afford various modalities of design workflows (forward, inverse, and hybrid) and support complex design tasks that require different levels of accuracy (in descending order: modularization, parametric design, and exploration), which are exemplified by three design examples derived with our CAD tool prototype. The contributions of this work include:

1. **A simulation method** that combines FEA and ML to simulate morphing materials fast and accurately.

2. **An ML architecture based on graph convolutional network (GCN)** that is adapted to topological morphing material systems.

3. **an exemplary simulator development pipeline for 2x2 grid structures** that comprises data generation, model training, and CAD toolmaking

4. **a CAD tool prototype and design examples** that demonstrate the enabled design space.

## RELATED WORK
### Simulation in Morphing Materials
Geometrical abstraction-based simulators are often used in morphing materials design, and trade physical accuracy for fast computation. While the prediction results can visualize the transformation trend, they are not sufficiently accurate to support design tasks that require high precision like modularization. In relatively small scales, Thermorph [1], Printed Paper Actuator [39], A-line [40], and bioLogic [46] combined parametric geometries with forward kinematics to simulate tree-topological patterns, but this approach is incompatible with more complex or larger patterns like 4DMesh [41] due to their omission of physical forces. To tackle more complex patterns, [32] and Geodesy [12,32] used linear mass-spring models to approximate the materials' transformation. Still, this approach requires taking small time steps to avoid divergence, leading to long simulation rollout (i.e., a trial of simulation) time and cannot afford real-time CAD interactions and iterations. Similarly, although elastic rods [31] have been used to assist the design of deformable objects, their limitations (i.e., tradeoff between noncircular cross-section shapes or viscoelastic materials [5]) make them inapplicable to certain morphing materials design spaces (e.g., the viscoelastic transformation of [40,41,47]). Compared to these methods, SimuLearn can provide more accurate predictions and support larger design spaces while requiring similar or less computation time.

Numerical methods like FEA have also been applied to predict material transformations [8] and are often used in standard commercial systems. These methods use physically-based material models and boundary conditions to produce more accurate results, and their accuracy allows for utility beyond visualization. For instance, FEA has been applied to design adaptive actuators [6], multi-stage transformations [6,7], and self-folding structures of complex topologies [50]. A recent work [47] also demonstrated using FEA as a backend engine to design robust artifacts made of composite materials. Yet, FEA involves establishing and solving large linear systems, making them time-consuming to perform even on supercomputing servers. Alternatively, Transformative Appetite [42] produced fast simulations by geometrically interpolating between precomputed FEA results, but this approach can only support a limited number of design parameters. Model reduction methods have also been used to achieve interactive FEA in animation [2] or material design [44]. Although they afford two-orders faster speeds, they also require precomputing the model's input motion and material modes, which leads to a delay when launching the editor. Changing the model's shape also requires reprocessing, thus making them less practical to use in the early stages of design where geometrical modifications are frequent. By contrast, SimuLearn uses abstract graphs to flexibly represent shapes that follow a specific topology and can take on more design variables while requiring little precomputation, leading to three-orders faster acceleration, larger design spaces, and better interactivity.

**Data-Driven Simulation**

Data-driven simulation methods have recently been used to accelerate simulation in various ways, such as numerical coarsening [10], subspace dynamics modeling [13], and reaction-diffusion [21]. These approaches use accurate simulators to trade precomputation effort for better runtime performance. When combined with ML, data-driven methods can also make simulations more accessible in various domains like fluid dynamics [20], biomechanics [22,24], and solid mechanics [29]. While ML-based techniques require additional data collection, and their generalizability are limited by the dataset, they also offer unique advantages such as parallelizability, end-to-end differentiability [49], and often three-orders faster speed. SimuLearn takes an identical approach and uses FEA as the source of data to ensure simulation accuracy. Moreover, in order to support the object-oriented modeling (i.e., constructing design by compositing elements) of morphing materials design, we take inspiration from the GCN in [3,26,34] and use graphical representations in this work. Unlike convolutional neural networks (CNN) [27] that require high-resolution voxelization/pixelization, GCN also takes advantage of the model's intrinsic topology to represent them with fewer yet more effective features and make ML models easier to train.

**Functional Simulation in HCI**

Simulations have been widely used in HCI to make inverse design tools for various material types. For elastic materials, [9, 23,48] used variations of FEA to enable users to predict and design shape-changing interfaces with complex deformation behaviors. For rigid materials, Forte [11], AutoConnect [19], and [36,45] used physical simulations to augment design tools and produce structurally optimized objects. In architectural scales, TrussFab [18], and TrussFormer [17] also used interactive simulation to guide users to design pavilions that met structural demands.

Other than design optimization, simulations also played a central role in computational fabrication. For instance, using simulation as a backend engine, Ion et al. [16] enabled users to create complex Metamaterial Mechanisms [15], [35,43] can optimally embed electronic components into 3D printed objects, and AutoConnect [19] empowered users to create 3D printable and robust connectors. Sequential Support [28] also used simulations to harness time-dependent material dissolution as a fabrication strategy. Situated among this literature, we believe that SimuLearn's speed and accuracy will allow available CAD tools to become more augmentative and effective in forward and inverse design tasks. Taking inspiration from Dream Lens [25], we also believe that SimuLearn can support generative tasks and allow users and computers to co-design morphing materials.

## OVERVIEW

**Material System**

Our 4D printing material system is based on polylactic acid (PLA) and is identical to the bending-based printing strategy of 4DMesh [41] (Figure 2). However, we constrain the grids to have a 2-cell by 2-cell configuration to simplify the ML problem space, and we opt to not use even smaller grids (i.e., 1x1 grids, rectangles) due to their confined design space. While the length of the beams may vary, their width and thickness are set at 7.2 mm and 4 mm, respectively. The actuators are quarterly assigned to the beams (Figure 2C), and the maximum curvature was measured to be 1.95 degrees/mm. We also make several improvements to 4DMesh's printing toolpath to facilitate FEA modeling (Figure 2A), which includes substituting the porous passive with solid (i.e., 100% infilled) constraint blocks and printing the joints as with alternating infill directions to minimize their transformation.

Printed structures are fixed on an aluminum frame to remain still and submerged throughout actuation in an 80 °C water bath (Figure 2B). Note that the grids are glued to the aluminum stand at the central joint, corresponding to FEA's fixed-joint assignments. Actuated grids are retrieved from water when the temperature drops below 60 °C, PLA's resolidification temperature. In our batch-to-batch printing and actuation consistency tests, we observe that a 150x150 $mm^2$ grid takes 45 minutes to print, and the diagonal span of grids may vary by 4.09% (with respect to grid dimension) after actuation. This number is regarded as the baseline accuracy requirement of SimuLearn.

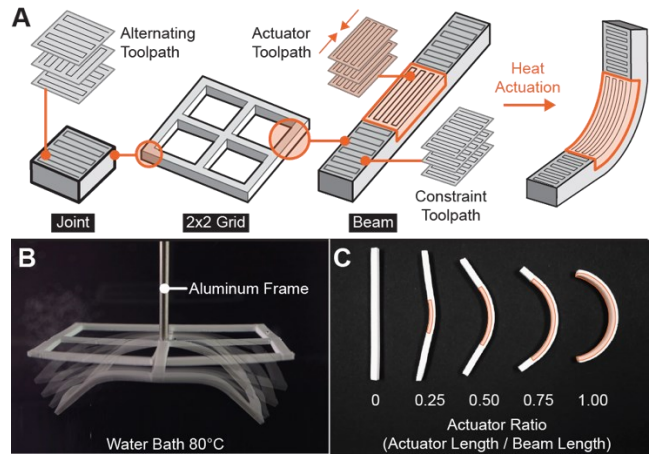

Figure 2. Our 4D printing material system - (A) Grid structure and toolpath design, (B) actuation setup, and (C) quarterly assigned actuators (printed and actuated). Actuators are highlighted with an orange outline in (A) and (C).

**Algorithm Design**

SimuLearn's implementation comprises two steps - dataset curation and ML model training. Dataset curation uses a physically-based FEA model to generate raw FEA results, which is later extracted to create a dataset for ML model training. Next, a GCN-based ML model learns from the dataset to become a generalized and accelerated simulator, which can then be used to compose design tools. In particular, SimuLearn relies on multilayer perceptron (MLP)-based GCN models to carry out fast computations.

This ML model allows us to represent the design using coarse elements described with succinct yet critical features, which drastically cuts down the number of computational units (Figure 3). Leveraging MLPs as nonlinear regressors, SimuLearn can also simulate with large time steps without compromising accuracy. Moreover, MLPs and GCNs are based on rapid, vectorized computations, making SimuLearn faster to compute than FEA and even comparable with geometrical methods.

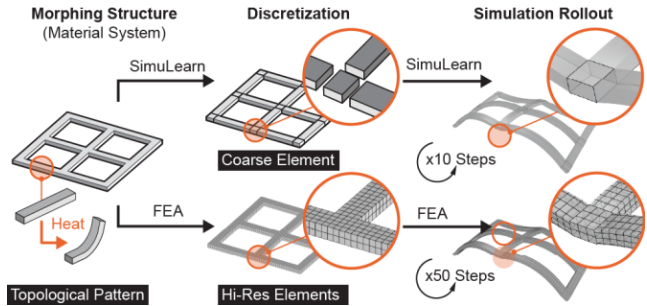

**Figure 3. Differences between SimuLearn and FEA.**

Figure 1A summarizes the computational theme of SimuLearn. Given an input design, we decompose the model into coarse elements represented by numeric features, compute pairwise interactions and elementwise updates with MLPs, integrate the update into the numeric features to derive the elements' status at the next time step, and repeat these steps until the simulation converges (i.e., no further transformation). In this computational flow, each iteration of the steps is identical to making one simulation increment in FEA. We can also arrange multiple SimuLearn engines in sequence to tackle complex physical systems that involve multiple stages - such as the sequential transformation of 4D printed PLA due to stress-release and creeping (Figure 6A).

**Design Tool and Workflows**

We incorporate the trained SimuLearn model in a design tool to demonstrate the forward, hybrid, and inverse design workflows supported by a fast and accurate simulator. A forward design workflow allows users to iteratively modify and simulate the model until satisfaction (Figure 1B, 11), enabling them to explore design options with low latency and without a clear goal in mind. On the other hand, an inverse design workflow (Figure 1C) helps users to achieve transformation goals when target shapes are identified. A hybrid workflow lies in between these two design modes - it allows the design tool to automate the objective aspects (e.g., optimizing design parameters towards a target shape) of the design process while enabling the users to enforce their subjective values (e.g., aesthetic concerns).

**IMPLEMENTATION DETAILS**

**Dataset Curation**

*FEA Modeling*
We use the analysis software Abaqus and follow [47] to establish a physically-based FEA model for our material system. This FEA model adopts a two-stage strategy to simulate 4D printed PLA: the first stage corresponds to the residual stress-induced transformation, and the second stage depicts PLA's creeping under gravity. The accuracy of this FEA model is reported to be above 95%. We refer readers to [47] for more technical details. The FEA solvers are configured to output a smooth animation of transformation processes - the first stage solver outputs ten equally spaced frames by procedurally releasing 10% of the total residual stress. In contrast, the second stage solver outputs only one frame due to relatively small deformations. Lastly, we use Abaqus2Matlab [30] to convert FEA results into .csv files.

*Data Generation*
We use a parametric script to generate different grid designs and FEA input files. The script initializes a design as a regular 2x2 grid and varies its morphing behaviors by randomly moving vertex positions in-plane and assigning actuators (Figure 4A). As a result, the generated grids would have different shapes and transformation behaviors while being topologically consistent, allowing for using regular expressions during feature extraction. It is worth noting that the variance of the design parameters bounds the ML model's generalizability, and if the ML model is presented with out-of-range grid design parameters, it is likely to produce less accurate results. Thus, in order to produce a simulator for a targeted design space, these factors should be taken into consideration and conveyed in the design tool.

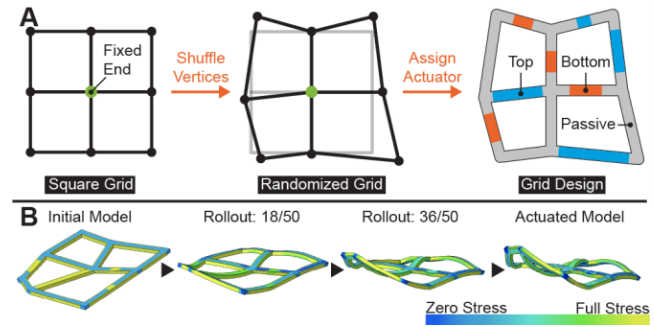

**Figure 4. (A) Random grid design generation procedure. (B) FEA result of a randomly generated grid.**

*Feature Extraction*
Computed FEA trials are used for feature extraction to obtain the training dataset. We rotate and mirror the simulation trials in-plane to eliminate orientational biases and procure more data points. At each timestep, a grid is represented as an abstract graph G = (A, N, E), in which the adjacency matrix A = $\{A_{ij}\}_{i=1...Ne,\ j=1...Nn}$ describes the connectivity between joints and beams, and the node and edge feature matrices N = $\{N_i\}_{i=1...Nn}$ and E = $\{E_i\}_{i=1...Ne}$ encode the joints' and beams' feature vectors, respectively.

Each edge feature vector $E_i$ encodes the information of three cross-sections located at the start, center, and end of a beam. The coordinates of the four corner vertices describe the shape of a cross-section, and the residual stress is represented by the stress field located at the Gaussian quadratures [33] around each of the corners (Figure 5A). On the other hand,

$N_i$ uses eight corner vertices to encode a joint's cuboid shape and omits the stress field information due to the lack of active transformation (Figure 5B). In addition to the physical information, $N_i$ and $E_i$ also have additional feature values to describe their design (i.e., a float value to indicate beam actuator assignments and a binary value to indicate fixed-end conditions of joints) and relative position to the fixed-end (i.e., the element's center point). Lastly, for each adjacent joint-beam pair, $A_{ij}$ uses a non-zero number to encode their face-to-face adjacency mode (Figure 5C).

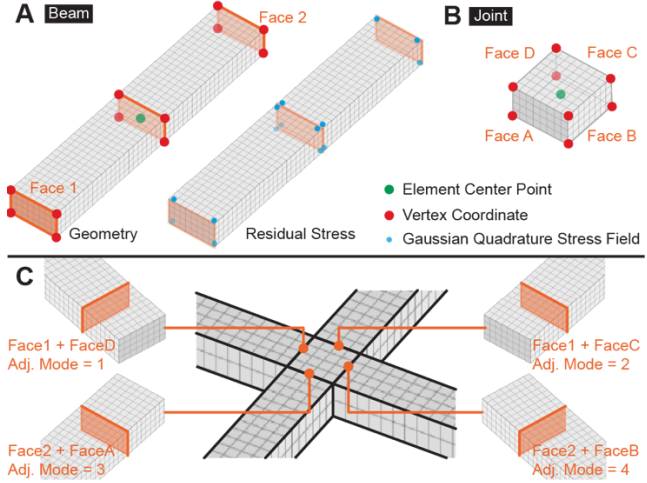

**Figure 5.** The feature sampling points of (A)beams and (B) joints. (C) An illustration of different adjacency modes.

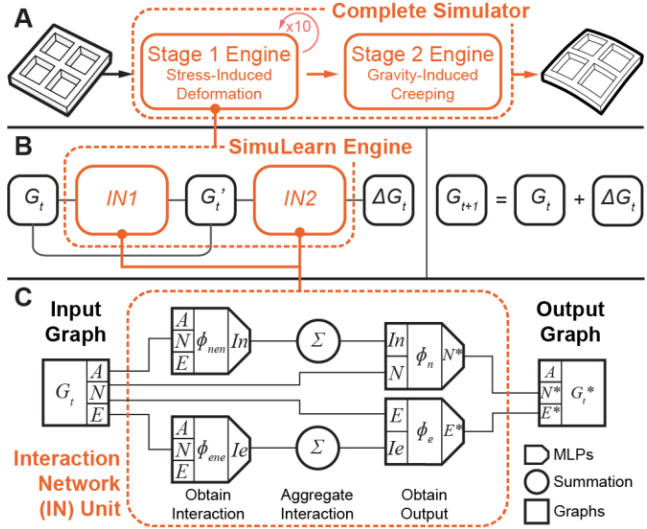

**Figure 6.** The hierarchy of our ML model architecture. (A) Using two SimuLearn engines to approximate the two-stage FEA model. (B) The double IN architecture of a SimuLearn engine. (C) The MLP layout within an IN unit.

### Machine Learning Model

*Model Architecture*
Figure 6 provides a hierarchical overview of our ML model architecture. At the top level, the complete simulator consists of two SimuLearn engines that correspond to each stage of the FEA solver (Figure 6A). The first engine recursively updates the input grid ten times to capture the first FEA stage's incremental simulation, whereas the second engine only updates once. At the next level, taking inspiration from [34], each SimuLearn engine uses two sequentially arranged interaction networks (INs) [3] to compute a grid's update (Figure 6B). Given $G_t$, a grid's graphical representation at time $t$, the first IN allows for element-wise interactions to propagate throughout the grid by obtaining a latent graph $G_t'$ that abstractly describes the summed effect subjected by all other elements over a beam or joint's transformation. The second IN then takes the concatenation of $G_t$ and $G_t'$ to compute the input grid's update $\Delta G_t$. Finally, the graph at the next timestep $G_{t+1}$ is obtained by adding $\Delta G_t$ to $G_t$.

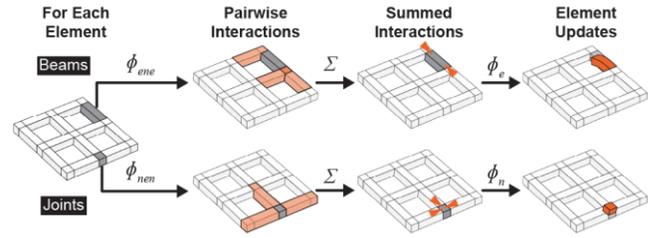

**Figure 7.** Visualization of GN units' forward computation.

**Algorithm 1. Interaction Network, IN**

**Input:** Graph, $G = (A, N, E)$
**for** each $A_{ij} \neq 0 \in A$ **do** // edge-node-edge interaction
    **for** each $A_{kj}, k \neq i \neq 0 \in A$ **do**
        Gather interaction pair $E_i$, $N_j$, $E_k$, $A_{ij}$, $A_{kj}$
        Compute interaction $Ie_{ik} = \phi_{ene}(E_i, N_j, E_k, A_{ij}, A_{kj})$
**for** each $A_{ji} \neq 0 \in A$ **do** // node-edge-node interaction
    **for** each $A_{jk}, k \neq i \neq 0 \in A$ **do**
        Gather interaction pair $N_i$, $E_j$, $N_k$, $A_{ji}$, $A_{jk}$
        Compute interaction $Ie_{ik} = \phi_{nen}(N_i, E_j, N_k, A_{ji}, A_{jk})$
**for** each node $N_i \in N$ **do** // node update
    Aggregate $In_i = \Sigma_j In_{rj}$ per receiver
    Compute output, $N_i^* = \phi_n(N_i, In_i)$
**for** each edge $E_c \in E$ **do** // edge update
    Aggregate $Ie_i = \Sigma_j Ie_{rj}$ per receiver
    Compute output, $E_i^* = \phi_e(E_i, Ie_i)$
**Output:** Graph, $G^* = (A, N^*, E^*)$

$\phi_{nen}$ : node-edge-node interaction network
$\phi_{ene}$ : edge-node-edge interaction network
$\phi_n$ : node output network
$\phi_e$ : edge output network
$In$ : latent node interaction vector
$Ie$ : latent edge interaction vector

INs are the fundamental building blocks of SimuLearn. Figure 6C characterizes an IN's forward computation to obtain its output. First, the model uses two interaction MLPs ($\phi_{nen}$ for node-edge-node and $\phi_{ene}$ for edge-node-edge interactions) to compute the pairwise interaction vectors ($In$ or *Ie* whose length is a hyperparameter), which describes a neighbor's influence over a receiver element. Next, for each element in the grid, the IN sums the interaction vectors that the element is the receiver of to obtain a convoluted interaction vector ($In_{conv}$ or $Ie_{conv}$) that represents the entire grid's influence over its transformation. Lastly, the element's corresponding output MLP ($\phi_n$ or $\phi_e$) then takes the element's feature and convoluted interaction vector to obtain their output ($N^*$ or $E^*$). Figure 7 visualizes this computation flow, and Algorithm 1 is a snippet of the forward computation of an IN. In an interaction pair, the first three items ($N_i$, $E_j$, $N_k$ or $E_i$, $N_j$, $E_k$) describe the sender, conduit, and receiver of interaction, and the last two items ($A_{ij}$, $A_{kj}$ or $A_{ji}$, $A_{jk}$) indicate the adjacency mode between the elements.

*Unsupervised Data Normalization*
In order to reduce redundant data variance and to improve feature quality, we statistically analyze the training dataset to produce normalizers for each MLP in our ML model. A normalizer applies a series of transformations to a data point to form the MLP's input, including moving the interaction pairs or elements to the spatial origin to remove locational variance, using principal component analysis (PCA) to reduce feature dimensions, and using affine transformation to produce zero-mean, unit-variance inputs for MLPs. In our implementation, setting the PCA information cut-off to 98% leads to halving the feature lengths, enabling faster model convergence and reducing overfitting. Note that all latent vectors are omitted during normalization.

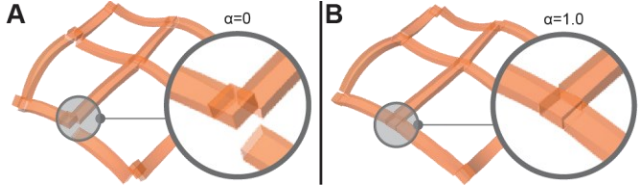

**Figure 8.** Rollout results at t=10 predicted by (A) the baseline model and (B) a model trained with the dislocation penalty.

*Loss Function*
When training the model with mean squared loss (MSE) alone, the vertices located at the junction of joints and beams are likely to become dislocated (i.e., becoming separated) after simulation, which violates the grid's topology and yields visually confusing results (Figure 8A). In, we add a penalty term to our objective function to constraint the model from producing vertex dislocation (Figure 8B):

$$Loss = L_{reg} + a \cdot (L_{disloc}), \text{ where}$$
$$L_{reg} = \left(\sum_i \|\widehat{N_i} - N^*\| + \sum_i \|\widehat{E_i} - E^*\|\right),$$
$$L_{disloc} = \sum_{(i,j) \in P} \|V_i - V_j\|_2$$

The first term $L_{reg}$ is the MSE between the model output $G^* = (A, N^*, E^*)$ and the FEA ground truth $\widehat{G} = (\widehat{N}, \widehat{E})$, and the second term $L_{disloc}$ the penalty term that measures the summed vertex dislocation. $V=\{V_i\}_{i=1...N}$ is the set of vertices encoded in $N^*$ and $E^*$, and $P=\{(i,j)\}_{i, j=1...N, i \neq j}$ indexes supposedly contiguous point pairs in $V$. Lastly, $a$ is the penalty strength and is regarded as a hyperparameter.

*Model Training*
Each MLP in our model comprises five hidden layers of logarithmically decreasing widths (e.g., 2048, 1024, 512, 256, 128). Since our model contains multiple MLPs and latent features, we train a SimuLearn engine as a deep network using batch gradient descent and an Adam optimizer. To improve the model's resilience against accumulated errors during simulation rollout, noises are added to the input graphs during model training:

$$Noise = (G_t - G_0) \cdot \mathcal{N}(0, \gamma^2)$$

$G_t$ and $G_0$ are a grid graphical representation at time t and 0, and $\mathcal{N}$ is a normal distribution with variance $\gamma$. In other words, the noise is proportional to the $G_t$'s cumulative update, and $\gamma$ defines the magnitude of the noise. Lastly, the hyperparameters of our method, $a$, $\gamma$, and dataset size, are determined with a hyperparameter grid-search (Figure 9). The optimal setting is identified as $(a, \gamma) = (1.0, 0.1)$ for their small dislocation error.

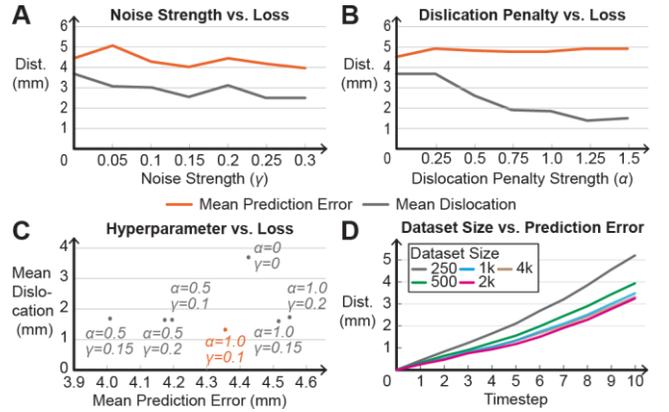

**Figure 9.** Hyperparameter search results for (A) noise strength and (B) dislocation penalty, (C) selected hyperparameter combinations, and (D) dataset size.

## RESULTS
We use the data generator to obtain 4,377 2x2 grid FEA trials. Depending on the grid size, each trial takes 8 to 14 minutes to compute (mean: 10.35 min.) on a consumer-grade desktop PC (8 core Intel i9-9900k processor at 5Ghz). The mean grid dimension (the largest span from the fixed-end to an outlying joint) is 94.44 mm (3rd and 96th percentile: 65.64 and 124.39 mm), and the average beam length is 51.29 mm (3rd percentile: 23.11 mm, 97th percentile: 80.19 mm).

**Performance Evaluation**
We benchmark a simulator with 2,000 randomly drawn FEA trials (1,600 for training, 400 as held-out test data). On average, a single rollout takes 0.61 seconds to complete (including input formatting, simulation, and writing result files), which is 1018x faster than using FEA on the same machine. SimuLearn also supports parallel, near real-time simulation using a GPU (1.94 seconds for 100 grids on an

Nvidia RTX 2080Ti), which is difficult to achieve with FEA due to the sheer computational cost. When measuring vertex coordinate errors between SimuLearn's predictions and FEA ground truths, the mean error is identified as 2.89 mm across all test data (Figure 10), which is 3.03% with respect to the dimension of grids (97$_{th}$ percentile: 6.93mm, 4.13%).

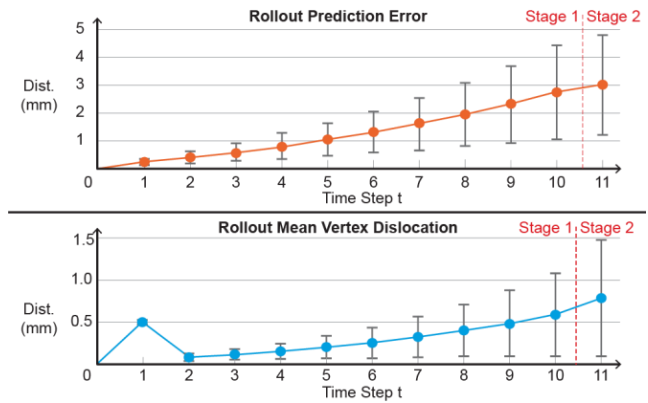

**Figure 10. Simulation rollout accuracy of 400 held-out data.**

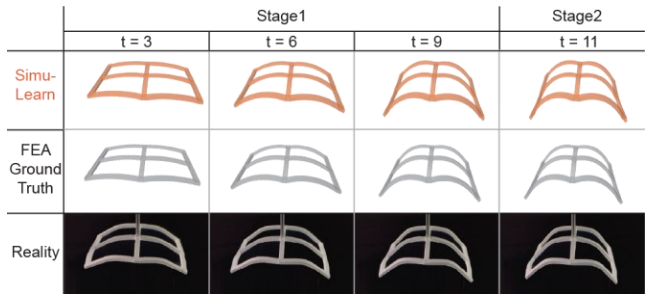

**Figure 11. Side-by-side comparison of SimuLearn, FEA, and physical ground truth. Grid size: 132.36 mm * 77.13 mm.**

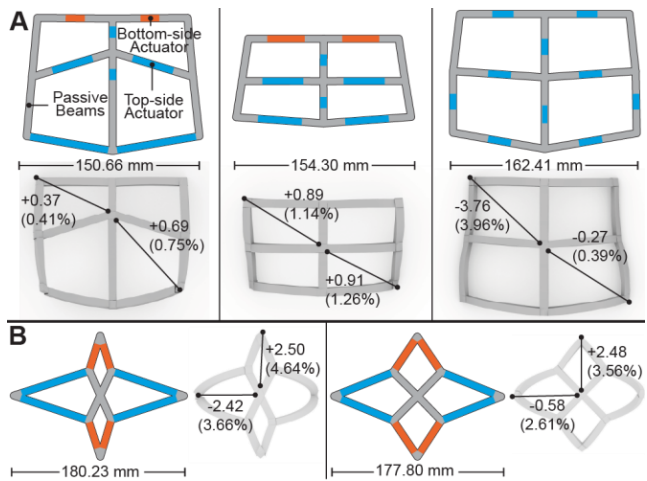

**Figure 12. SimuLearn accuracy versus real grids shown in the (A) lampshade and (B) aggregated table design. (units: mm)**

Compared to physical prototyping, SimuLearn allows users to preview a grid's transformation x9000 faster (90 minutes for printing and triggering the grid shown in Figure 11). As for accuracy, since vertex coordinates are unavailable, we measure the distances between several feature point pairs and report the mean error to be 2.22% for the grids shown in Design Examples (Figure 12), anecdotally implying a 97.78% accuracy with respect to the physical truth. While this number is inconclusive due to the limited number of samples, the error is lower than the fabrication error. Thus, it is sufficiently accurate to support design tools and tasks. In terms of smoothness, Figure 1C and 16B showed small changes in design parameters would not lead to drastic changes in simulation results. A more comprehensive experiment is also provided in the Supplementary Materials.

**Design Tool Implementation and Supported Workflows**

*Forward Workflow*

The design tool is implemented as a Rhinoceros 3D and grasshopper script, such that users can model and simulate the grids in a single environment and generate fabrication files. When forward-designing a 2x2 grid, users can initialize its shape by choosing from a predefined library or by drawing the grid's skeleton as polylines (Figure 13A) and assigning actuators to beams (Figure 13B). Simultaneously, a validation subroutine will check the design against topological constraints imposed by the material system and the dataset to ensure its compatibility with the simulator (Figure 15). Once validated, users can then use SimuLearn to predict the grids' transformation and navigate between each timestep with a slide bar (Figure 13C). Users can make design decisions and manual iterations based on the simulation results, but the design tool also has a set of functions to assist users in achieving desired transformations (Figure 14). Once completed, the tool then processes the model design into G-code files for fabrication (Figure 13D).

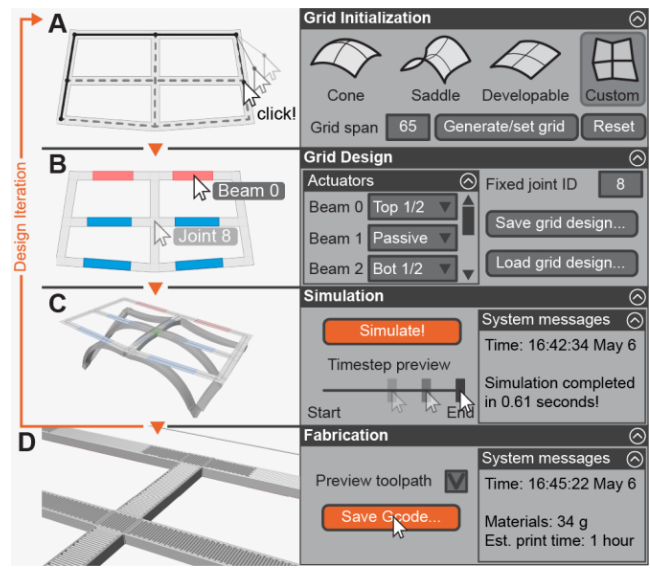

**Figure 13. A forward design workflow - (A) initializing a grid by sketching its skeleton, (B) assigning actuators and fixed joints, (C) simulating transformation, and (D) export print files.**

*Inverse and Hybrid Workflow*

In an inverse or hybrid design workflow, following the initialization of grid design, the user can specify vertex transformation goals as target points to the design tool

(Figure 14A). The design tool then modifies each of the parameters (i.e., changing beam actuator assignments by ± 0.25 or moving joint positions along octagonal directions with a specified distance) to generate design variations, batch-simulates their transformations, and compares the results against the target points to rank the effectiveness of design modifications. The rankings are determined by the averaged distance between target point pairs. In a hybrid design workflow, the top-five modifications are presented to the user to choose from (Figure 14B). In contrast, the top-ranked update is automatically applied (Figure 1C) in an inverse design workflow. These steps can be repeated for as many times as the user specifies, and each epoch can be completed in near real-time (2 seconds for simulation and ranking, 8 seconds for rendering the interface). Qualitatively speaking, this gradient-free, brute-force method provides a simple yet effective way to perform design optimization.

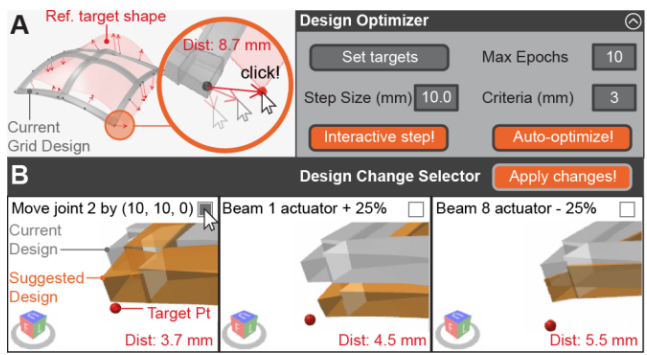

Figure 14. Inverse and hybrid design workflows. (A) User specifying transformation goals in the design tool. (B) The design tool suggesting ranked design modifications for the user.

*Design Validation*

During the modeling step, the validation subroutine provides visual cues to guide users to design grids that comply with the material system's intrinsic topology. The design tool presents two types of messages to users: *errors* (Figure 15A, B) that make the grid topologically incompatible with the simulator and *warnings* (Figure 15C, D) that may affect simulation accuracy. From a user's perspective, error messages will block the simulator from running until addressed, whereas warning messages will only prompt users to modify but do not hinder simulation.

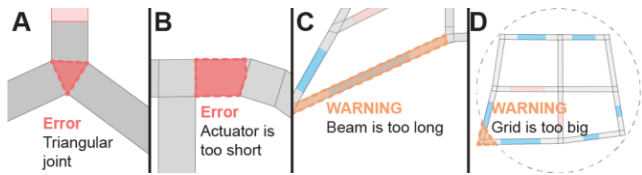

Figure 15. Validation messages showing (A) joint configuration error, (B) actuator length error, (C) beam length warning, and (D) grid size warning.

### DESIGN EXAMPLES

We use three design examples to demonstrate the workflows enabled by SimuLearn - inverse design of a lampshade, hybrid design of an aggregated table, and forward design of a decorative joinery. The lampshade example demands the highest of simulation accuracy and smoothness due to the optimization task, whereas the joinery requires the least as simulations are only used to visualize transformations.

**Inverse Design Workflow: Modularized Lampshade**

SimuLearn's accuracy affords design tasks that require high precision, such as patching surfaces to form larger structures [38]. In this example, the user first creates a surface model of the lampshade, but its large dimension makes it difficult to 4D print as a whole. Thus, the designer patches the surface with three repeating modules to make it more fabricable. Each module is fitted with a 2x2 grid (Figure 16A) by specifying target points for vertices, and the design is carried out using the design tool's inverse design function (Figure 16B). Once optimized, the user then manually reorients the modules back to the surface model to generate an assembly preview. The design tool makes 22 iterations and explores 1,958 design variations in 12 minutes to bring the mean fitting error to 4.44 mm (i.e., the distance between target point pairs). Note that most of the computation time is used to render results into the design interface and the actual simulation time is less than 50 seconds.

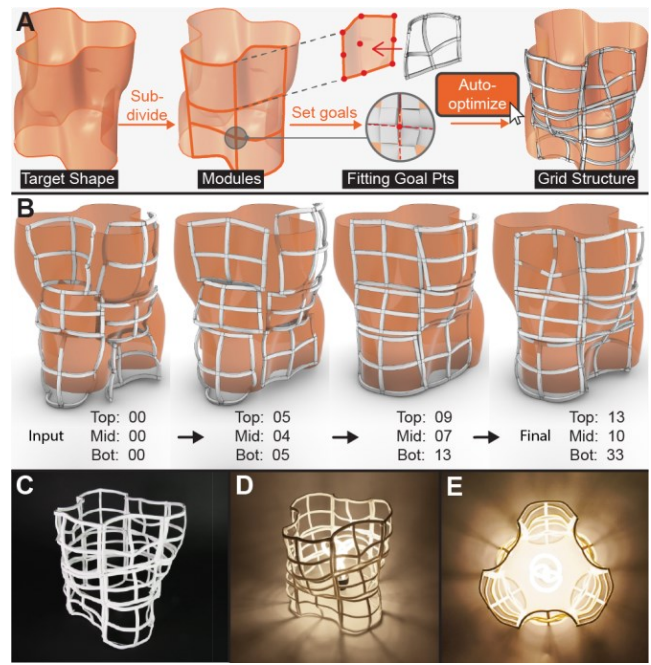

Figure 16. Lampshade design - (A) design scheme, (B) selected optimization epochs (epochs labeled at the bottom), and (C) assembled and (D, E) illuminated lampshade.

Figure 16C-E shows the printed, actuated, and assembled lampshade design. When comparing SimuLearn results with the physical reality, the max errors are 3.95 mm (4.21%), 2.422 mm (3.30%), and 3.572 mm (3.57%) for the top, middle, and bottom piece, respectively. The errors are sufficiently small and the modules are assembled without any noticeable issue. We report that the target shape is unachievable using previous methods because its geometry violates 4DMesh's [41] algorithmic constraints. The folding-

or wire-based strategy of [1,40] also cannot produce artifacts with sufficient structural strength. More, modularization also helps to compartmentalize printing time and material usage, thus helps to mitigate fabrication risks. This design also shows that a fast and accurate simulator like SimuLearn can help us produce larger-scaled 4D printing structures, further advancing the fabrication flexibility of 4D printing.

design often requires high interactivity, and this modality of morphing materials design is only achievable with SimuLearn as a back-end engine. Other simulation methods will either make the design workflow impractically slow or lack the accuracy needed by parametric design schemes.

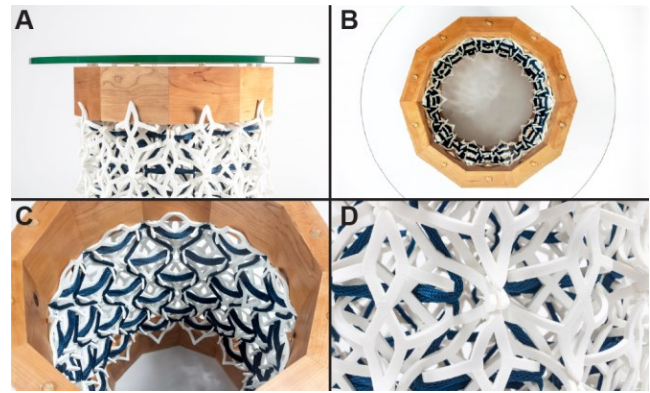

Figure 18. Aggregated table (A) side, (B) top, and (C) inside view pictures, and (D) the weaving technique detail.

The final design is achieved after seven iterations over 15 minutes, in which only less than 1 minute is used for simulations. The max error between SimuLearn predictions and physical prototypes are 5.22 mm (8.02%) and 3.96 mm (5.95%) for the grids in the unit. The table is assembled using a Native American off-loom bead weaving technique (Figure 18). The level of detail and structural overhangs make this design difficult to fabricate with conventional 3D printing methods, and it would also be uneconomical to print using dissolvable support materials. In total, sixty cell-units are used in to produce a 52.6 cm tall, 46.8 cm wide structure.

**Forward Design Workflow: Decorative Wood Joinery**
In this example, the designer is tasked to create a 4D printed diagonal support for a miter wood joint. The designer adopts a forward design workflow by manually modeling the grids. The simulations are used to avoid collisions, identify insertion hole placements, and predict interlocking behaviors (Figure 19A) between units. SimuLearn's speed allows the designer to make quick iterations and explore various design options (Figure 19B). In total, the designer produces 4 design variations (4-8 iterations each) in 25 minutes (1.5 minutes for simulation and 22.5 minutes used for modeling).

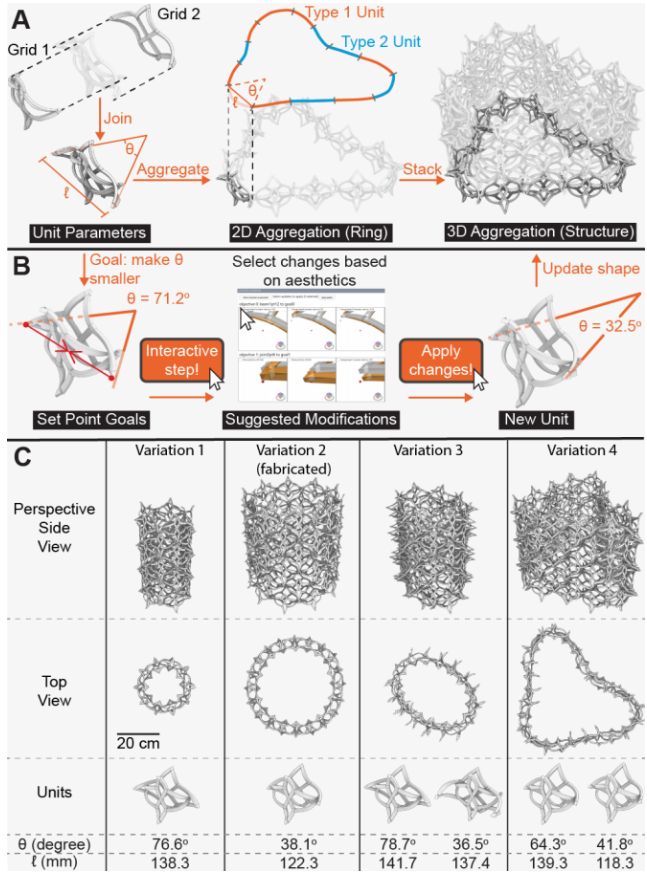

Figure 17. Aggregated table design - (A) parametric design scheme, (B) the hybrid design workflow to change the aggregation's shape, and (C) selected design variations.

**Hybrid Design Workflow: Aggregated Table**
This structure is created by connecting multiple cell-units to create a ring and stacking several rings to form the entire aggregation. A cell-unit is made of two intersecting 2x2 grids, and their tangential lines determine the contact angle θ of the unit, which consequently decides the curvature of the aggregation (Figure 17A). In this parametric scheme, the designer cannot directly control the aggregation's shape but have to indirectly change the cell-unit contact angle instead (Figure 17C). To do so, the designer uses the hybrid workflow to specify the design tool to bring two vertices closer or away from each other, then select from the ranked modifications to update the cell-unit. During this co-design process, the design tool suggests viable options based on the simulation results, and the designer makes aesthetic judgements to make sure the aggregation is aesthetically consistent throughout the structure (Figure 17B). Parametric

The final design (Figure 19C-E) comprises three 2x2 grids that interlock and fasten together by sequential actuation. The centerpiece is first actuated and assembled into the wood joint, then the two side pieces are actuated while being inserted into the wood panels' slits to fasten the structure together. When compared to the physical reality, the max simulation errors were 2.80 mm (8.44%) and 3.09 mm (7.90%) for the center and side pieces, respectively. Noticeably, due to the model's large size, the corner joint blocks appear distorted at the end of the simulation (Figure 19F), but the prediction result still captures the trend of transformation and an approximative shape of the actuated grid, thus satisfies the accuracy demand of this design task.

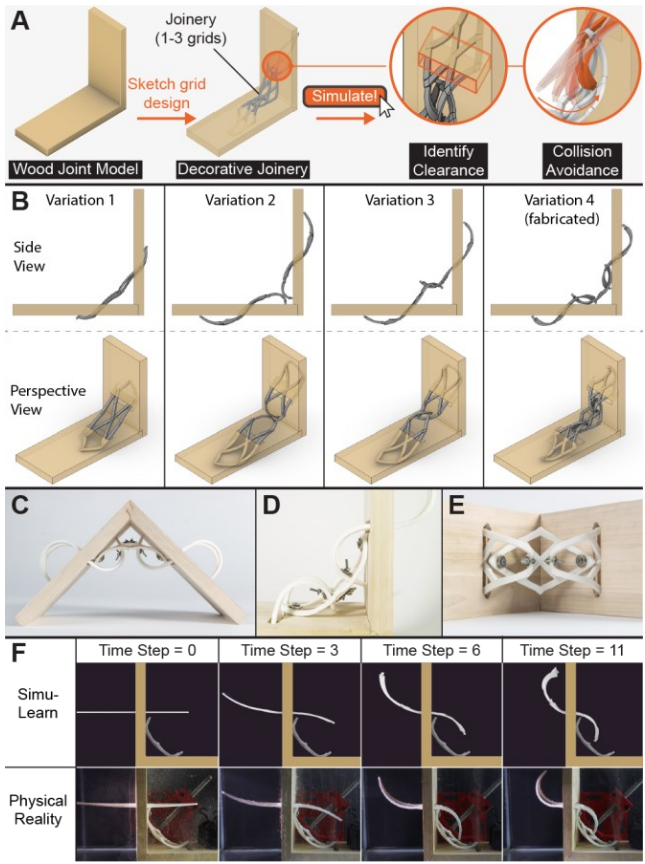

Figure 19. Wood joinery design - (A) Forward design scheme, (B) selected design variation, (C, D) side views and (E) details of the assembled design, and (F) transformation process.

## LIMITATIONS

### FEA Limitation

SimuLearn's accuracy is limited by the data source and is susceptible to the limitations of the FEA model. In this work, although our FEA model is physically accurate, it does not account for collisions during the transformation process. Although these phenomena are unlikely to occur in our material system due to their transformation capacity, future work may take inspiration from [3] to take account for collision and open new design spaces.

### ML Simulator Accuracy

While our physical prototyping results showed that SimuLearn's speed and accuracy could readily support and facilitate their design workflows and tasks, there is still room for performance improvements. For instance, our current method does not use the temporality of simulation trials to train the model. Inspired by [34], we speculate that adopting recurrent ML models may further improve the accuracy of SimuLearn. Incorporating other ML techniques such as encoder/decoder, system identification, or hierarchical convolution [26] may also lead to improved performances. Future works may also leverage our pipeline to generate larger datasets in order to mitigate the dimension issue observed in the decorative joinery design example.

### Development Cost

SimuLearn trades development time for workflow conveniences by using FEA to curate large datasets for training ML-based simulators. In this work, we prioritize our data generation for the design parameters that we deem most important. To incorporating new design parameters, developers would have to curate new datasets to update the simulator. Indeed, when targeting at a more general design space, methods that do not require training on any possible topology may appear to be more economical. Yet, the development cost of SimuLearn can also be easily justified by its three-orders faster workflow acceleration and parallelizability, especially when the design tool is mass-deployed or repeatedly used. We also believe that SimuLearn allows developers to compose augmentative design tools for well-established morphing material systems like 4D printing, thus contributing to the democratization of advanced fabrication technologies.

## FUTURE WORK

### Generalizability and Scalability

While this work is adapted to a specific material system, SimuLearn's algorithm is also adaptable to other material systems by exchanging the FEA model and/or the feature representation. E.g., SimuLearn can adapt to Geodesy [12] by describing the continuous shells as aggregations of rectangular patches, which are then represented by their corner points, or it can further adapt to Transformative Appetite [42] by swapping the FEA model from stress-release PLA to swelling gel. Existing works have also validated the viability of ML-based physics in various engineering and design contexts [49].

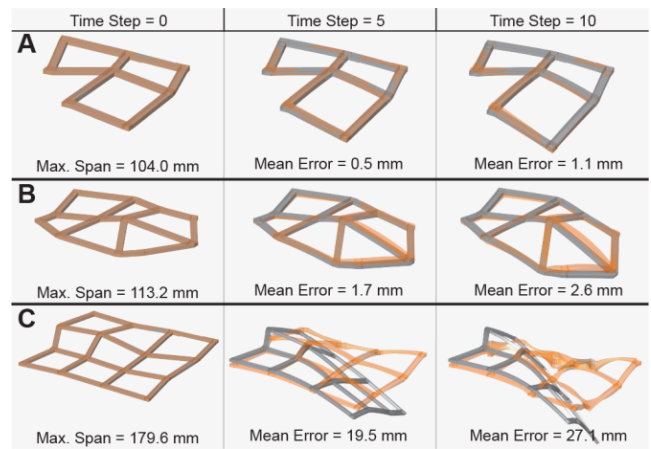

Figure 20. Simulation results of topologically mutated grids - (A) a 2x2 grid with partial removal, (B) a 2x3 grid, and (C) a 3x3 grid. (orange: SimuLearn result, grey: FEA ground truth).

As for scalability, GCNs intrinsically generalize to designs that have different numbers of units and are adaptable to different length scales [3], and the only limitation is dataset coverage. Anecdotally, we observe that the simulator can generalize to unseen grid topologies (i.e., having missing or extra beams) if their geometrical dimensions are within the dataset's coverage (Figure 20). Nevertheless, SimuLearn can also be trained to tackle topologically larger grids (e.g., a 4x4 grid) by expanding the dataset to cover targeted topologies and increasing the degree of convolution in the ML architecture. Note that the computation speed would remain identical because the elementwise transformations can be computed in parallel. We speculate that while adapting SimuLearn to larger grids would quadratically scale up the parameter space (i.e., elements may locate further from the fixed joint and be subjected to higher magnitudes of stresses), the amount of element data available for training MLPs would also increase quadratically. Thus, it may be possible to achieve an identical accuracy using the same amount of FEA trials — though further research is necessary in order to validate this conjecture. Nonetheless, we argue that while the simulator is limited to 2x2 grids, its speed and accuracy affords users to design larger structures using a modularization approach with even higher efficiency than previous work [41].

**SimuLearn-Based Design Agents**
Currently, the inverse design function optimizes the model with an unguided brute-force approach. Future works may consider using different optimization approaches to achieve better results. In particular, SimuLearn's parallelizability and speed lend itself well to genetic algorithms and evolutionary computing that require frequent performance evaluations. More than being faster, SimuLearn also enables converting indifferentiable simulations like FEA into differentiable computations, which can be leveraged to create gradient-based optimizers. Similar methods have also been shown in robotics for efficient control policy-finding [4] and co-design [14]. Situating this concept in HCI, SimuLearn as a backend engine will allow CAD tools to simulate, evaluate, and suggest designs in real-time to inform high-quality decisions. With SimuLearn's debut, we also envision conversational design agents to emerge in the shape-changing interfaces and morphing materials context.

**CONCLUSION**
SimuLearn combines FEA and ML to enable physically accurate and real-time simulations for morphing materials. Results show that SimuLearn is nearly as accurate as state-of-the-art methods while being orders of magnitude faster. It also enables design tools to become multimodal platforms that support a broad spectrum of design workflows. Beyond the grid- and PLA-based material system presented in this paper, we also believe that SimuLearn can generalize to other topological patterns or morphing materials by swapping the representation and FEA model.

SimuLearn, as an enabling technology, is particularly well-suited for the HCI community. Not only because of its effectiveness in improving design efficiency, but also because its interactivity allows users and computers to co-design, paving the way for human-AI collaborations to unfold in the design field. We also believe that SimuLearn can augment morphing material CAD tools to become conversational, educative, and accessible to the public. As the HCI community accumulates growing interests toward harnessing active material behaviors, SimuLearn will likely enrich the available design and technology toolbox and empower us to unfold the potentials of active, smart, and morphable materials. With this vision, we seek to democratize SimuLearn by sharing its source codes at https://github.com/morphing-matter-lab/SimuLearn.

**ACKNOWLEDGEMENT**
This research was supported by the Carnegie Mellon University Manufacturing Future Initiative, made possible by the Richard King Mellon Foundation. We thank Angran Li, Daniel Cardoso Llach, Guanyun Wang, Ardavan Bidgoli, and Michael Rivera for providing feedback during the conceptualization stage. We would also like to acknowledge Jesse Gonzalez, Alex Cabrera, and the reviewers' comments that helped to improve the quality of this paper.